\documentclass[useAMS,usenatbib]{mn2e}
\usepackage{amssymb}
\usepackage{graphics}

\def\eqref#1{equation~(\ref{#1})}
\def\eqrefs#1#2{equations~(\ref{#1})-(\ref{#2})}
\def\apref#1{Appendix~\ref{#1}}
\def\scref#1{Section~\ref{#1}}

\def\arg{\mathop{\mathrm{arg}}\nolimits}
\def\sigmalevel#1{\ensuremath{#1}-\ensuremath{\sigma}}

\newcommand{\obs}{\mathrm{\rm obs}}
\newcommand{\TD}{T_{\rm D}}

\title[Periastron Precession of TEPs]{Periastron Precession Measurements
in Transiting Extrasolar Planetary Systems at the Level of General Relativity}
\author[A. P\'al and B. Kocsis]{Andr\'as P\'al%
$^{1,2}$\thanks{E-mail: apal@cfa.harvard.edu}
and Bence Kocsis$^{1,3}$\thanks{E-mail: bkocsis@cfa.harvard.edu} \\
$^{1}$Harvard-Smithsonian Center for Astrophysics,
	60 Garden street,
	Cambridge, MA, 02138, USA \\
$^{2}$Department of Astronomy, Lor\'and E\"otv\"os University,
	P\'azm\'any P. st. 1/A,
	Budapest H-1117, Hungary \\
$^{3}$Department of Atomic Physics, Institute of Physics, Lor\'and E\"otv\"os University,
	P\'azm\'any P. st. 1/A,
	Budapest H-1117, Hungary}

\begin{document}

\date{Accepted ..., Received ... ; in original form ...}

\pagerange{\pageref{firstpage}--\pageref{lastpage}} \pubyear{2008}

\maketitle

\label{firstpage}

\begin{abstract}
Transiting exoplanetary systems are surpassingly important among
the planetary systems since they provide the widest spectrum of
information for both the planet and the host star.
If a transiting planet is on an eccentric orbit,
the duration of transits $\TD$ is sensitive to
the orientation of the orbital ellipse relative to the line of sight.
The precession of the orbit results in a systematic variation
in both the duration of individual transit events and the
observed period between successive transits, $P_{\obs}$.
The periastron of the ellipse slowly precesses
due to general relativity and possibly the
presence of other planets in the system. This secular
precession can be detected through the long-term change in $P_{\obs}$
(transit timing variations, TTV) or in $\TD$ (transit duration variations, TDV).
We estimate the corresponding precession measurement precision for repeated
future observations of the
known eccentric transiting exoplanetary systems
(XO-3b, HD~147506b, GJ~436b and HD~17156b) using existing or planned
space-borne instruments. The TDV measurement improves the
precession detection sensitivity by orders of magnitude over the TTV measurement.
We find that TDV measurements over a $\sim4$\,year period can typically
detect the precession rate to a precision well exceeding the level predicted
by general relativity.
\end{abstract}



\section{Introduction}

Since the discovery of the first transiting extrasolar planet
\citep{charbonneau2000,brown2001}, the number of such systems
has increased to more than 30\footnote{See http://exoplanet.eu
for up to date information}.
These transiting extrasolar planets (TEPs) provide unique
information on the properties of the system.
Based on the geometry provided by the transit light curve(s),
the inclination, the physical radius and mass, therefore the density
and the surface gravity can be derived, in addition to the mass of the planet.
Moreover, the time between successive transits can be measured with
an exceedingly high accuracy ($\sim 10^{-6}$~--~$10^{-7}$, relative to the period).
The detection of long--term transit timing variations can be used to learn more
beyond the properties of the parent-star system \citep{miralda2002,steffen2007}. They can be
indicative of the presence of other planetary companions \citep[see e.g.][]{holman2005,agol2005,millerricci2008},
co-orbital companions \citep[Trojans, see][]{ford2007},
or satellites \citep{simon2007} in the system, could provide information on the
oblateness of the host star, or can be used to detect the additional prograde periastron
precession predicted by general relativity (GR) \citep{miralda2002,heyl2007}.
Secular variations in the semimajor axis (and therefore in the transit timing)
are also predicted on the time scale of stellar life due to the anisotropic
light redistribution \citep[a.k.a. Yarkovski-effect, see][]{fabrycky2008}.
Furthermore, \citet{iorio2006} has shown that TEP observations can in principle
also test the gravitoelectric correction of GR
by measuring the radial velocity amplitude and transiting periodicity simultaneously,
in order to verify that the third Kepler's law requires a semimajor axis
dependent correction.

In a pioneer study, \citet{miralda2002} derived the modification of the observed time period
between successive transits $P_{\obs}$, called transit timing variations (TTVs),
caused by the standard periastron precession due to GR \citep[e.g.][]{misner1973} and the perturbations of other planets if present.
Recently, \citet{heyl2007} have extended these studies and estimated the precision of
precession rate measurements for long--term mock observations of eccentric
transiting extrasolar planets (ETEPs). Both studies restricted to small eccentricities.
At that time, the existence of close eccentric planets was known
only through radial velocity measurements, and no ETEPs had been observed.
Since their publication, four transiting extrasolar planets have been discovered
with significant eccentricity: XO-3\lowercase{b} \citep{johnskrull2007},
HD~147506\lowercase{b} \citep[a.k.a. HAT-P-2,][]{bakos2007},
GJ~436\lowercase{b} \citep{gillon2007,butler2004},
and HD~17156\lowercase{b} \citep{fischer2007}.
Therefore it is now possible, for the first time, to make specific predictions
for future, long--term measurements of periastron precession effects for real
exoplanetary systems.

In this paper we determine the precision by which repeated long--term future
ETEP observations will be able to
detect the periastron precession rate for existing systems.
In addition to TTVs, i.e. the
slow modulation of $P_{\obs}$ considered previously \citep{miralda2002,heyl2007},
the periastron precession also changes the time durations $\TD$ of individual transits.
We examine whether these transit duration variations (TDVs) can be used to
improve the sensitivity of periastron precession measurements.
We estimate the precession rate measurement precision for long term
repeated observations of $P_{\obs}$ and $\TD$ for the known ETEPs. Since several of the
observed ETEPs have large eccentricities, we derive expressions for both TTVs and TDVs
which are applicable for arbitrary eccentricities.
We estimate whether future observations of currently known ETEPs
will be able to reach the sensitivity necessary to test the prediction of GR,
using existing or planned space-borne instruments.
We refer the reader to a recent independent study by
\citet{jordan2008}, of precession rates in eccentric transiting
extrasolar planets.

The next section of this paper introduces the geometrical description
which is the basis of our calculations, and derives the expected
transit timings and durations for planets orbiting a star with an
arbitrarily large eccentricity. In \S~\ref{s:realsystems}, we utilize our results for the
confirmed four ETEP systems, and give predictions
for future observations of periastron precession with space-borne observations.
Our conclusions are discussed in \S~\ref{s:summary}.


\section{Transit timings and durations for eccentric orbits}\label{s:transit}

\begin{figure}
\resizebox{8cm}{!}{\includegraphics{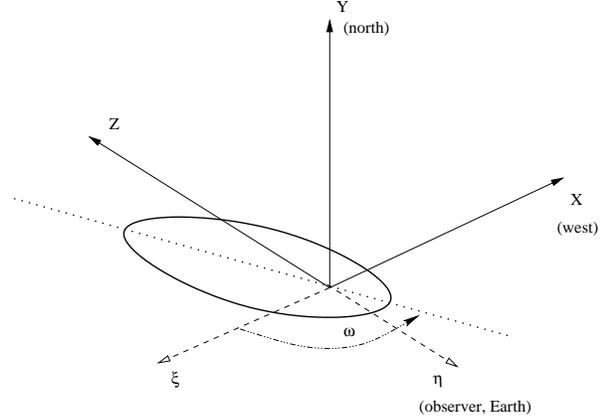}}
\caption{%
The geometry of the orbit of the transiting planet. The plane of the
sky is defined by the $X+$ (west) and $Y+$ (north) axes while the $Z+$
axis points away from the Earth. The orbital plane is defined by the
axes $\xi+$ and $\eta+$, where $\xi+$ is anti-parallel with $X+$ axis,
$\eta+$ is in the plane of $(Y+,Z+)$ and the angle between $Y+$ and $\eta+$
is the inclination, nearly $90^{\circ}$. The major axis of the orbit
is marked by the dotted line.}\label{fig:geometry}
\end{figure}

The reference frame used for the description of exoplanetary systems
as well as for binary/multiple stellar systems is fixed to the sky: the
plane of the sky is defined by the $(X+,Y+)$ while $Y+$ points towards
to north. For planetary transit observations, the line-of-sight lies
close to the orbital plane,
i.e. perpendicular to the plane of the sky.
The orbit is given by Cartesian coordinates $(\xi,\eta)$, where $\xi+$
is parallel to $X-$ and $\eta+$ oriented toward the observer (see also Fig.~\ref{fig:geometry}).
The Lagrange vector or eccentricity vector is given in these coordinates as
$(k,h)=(e\cos\omega,e\sin\omega)$.
Let us define the angle $\varphi_0$ as the angle relative to $\xi+$ in the orbital plane
at the instance\footnote{i.e. at the center of a transit} of the transit. From the definitions of $(\xi,\eta)$,
observing from Earth is equivalent to setting $\varphi_0=\pi/2$.

Now let us denote the mean longitude of the planet at the instance of the
transit by $\lambda$. For a circular orbit, $\lambda=\varphi_0$. From its standard
definition in celestial mechanics, it is straightforward to derive the mean longitude
for arbitrary eccentricities (see \apref{appendixlambdaderiv}). The result is
\begin{eqnarray}
\lambda & \equiv & \lambda(\varphi_0,k,h)\equiv \lambda(\varphi_0,e\cos\omega,e\sin\omega)= \nonumber \\
 & = & \arg\left(k+\cos\varphi_0+\frac{he_{\perp}}{2-\ell},
h+\sin\varphi_0-\frac{ke_{\perp}}{2-\ell}\right) - \nonumber \\
 & & -
\frac{e_{\perp}(1-\ell)}{1+e_{\parallel}}, \label{lambdaattransit}
\end{eqnarray}
where $e_{\parallel}=k\cos\varphi_0+h\sin\varphi_0$,
$e_{\perp}=k\sin\varphi_0-h\cos\varphi_0$ are the components of the eccentricity vector relative to the
line-of-sight, $\ell=1-\sqrt{1-e^2}$ is the oblateness of the orbit,
and $\arg(x,y)=\arctan(y/x)$ if $x\geq 0$ and $\pi+\arctan(y/x)$ otherwise.
Plugging in the observed values of $e$ and $\omega$,
\eqref{lambdaattransit} provides a simple way of calculating the mean longitude of the orbit
for an arbitrary transit observation. Note that this formalism omits the
direct usage of the mean anomaly, true anomaly and eccentric anomaly which
have no meanings for $e\to0$. All of our derived formulae are based
on the well--behaved parameters $\lambda$,
$k$, and $h$, and therefore are valid for arbitrary eccentricities.

In the following subsections, we derive the expressions describing TTVs and TDVs, discuss the corresponding
observational implications and estimate the precision of periastron precession observations.


\begin{table*}
\caption{Basic data of the four known eccentric transiting
exoplanetary systems: the mass ($M_\star$) of the parent star,
period ($P$, in days), eccentricity ($e$)
and the argument of pericenter ($\omega$),
the half-duration of a transit event ($H$, in days)
and the impact parameter ($b$). In the last two columns, we provide
the calculated values of the secular period caused by the GR periastron precession, and the minimum mass -- semimajor axis ratio
for a hypothetical exterior perturber at $a_2\gtrsim 3a$ that would lead to the same magnitude of periastron precession.
}\label{table:basicdata}.
\begin{tabular}{llrlllllr}
\hline
System	 	& $M_\star/M_{\odot}$	& $P$ (d)	& $e$			
		& $\omega$ (degrees) 	& $H$ (d)			
		& $b$
		& $P_{\rm sec}$ (years)	& $\frac{m_2/M_{\oplus}}{(a_2/3a)^3}$\\
\hline
HD~147506b	& $1.298\pm0.07$	& $ 5.63341 $	& $0.520\pm0.010$	
		& $179.3\pm3.6$		& $0.083$
		& $0$
		& $19790\pm740$ 	& $12.2$ \\

XO-3b		& $1.41\pm0.08$		& $ 3.19154$	& $0.260\pm0.017$	
		& $344.6\pm6.6$		& $0.050$
		& $0.8$
		& $9280\pm410$  	& $15.9$ \\

GJ~436b		& $0.452\pm0.013$	& $ 2.64385 $	& $0.150\pm0.012$	
		& $351.0\pm1.2$		& $0.065$
		& $0.85\pm0.02$
		& $15180\pm400$  	& $2.6$ \\

HD~17156b	& $1.2\pm0.1$		& $21.21725 $	& $0.6717\pm0.0027$	
		& $121.23\pm0.40$	& $0.098$
		& $0.50\pm0.12$
		& $143000\pm7900$	& $5.9$ \\
\hline
\end{tabular}
\end{table*}

\subsection{Modulation of the transit period}

The period between successive transits $P_{\obs}$ is modified by a possible slow precession of the orbital elements.
These modulations are referred to as transit timing variations (TTVs).
We derive the modulation of $P_{\rm obs}$ due to periastron precession in two steps.
First we demonstrate that the change in the period between successive transits is simply related to the change
in the mean longitude $\Delta \lambda$. Then relating the mean longitude shift to the periastron precession
rate we derive the expected TTV rate.

Let $P_0$ be the orbital period, $n=2\pi/P_0$ be the mean angular
velocity. The mean longitude of the planet increases steadily in time,
$\lambda=nt+\lambda_0$. The variation in the mean longitude at
the transit center would result in
a variation in the transit cadence. During a transit at time $t_1$,
the mean longitude is $\lambda_{\rm tr}(t_1)=nt_1+\lambda_0$,
and after an observed revolution, at the instance $t_2=t_1+P_{\obs}$
it is $\lambda_{\rm tr}(t_2)+2\pi=nt_2+\lambda_0$.
Therefore the observed period between transits is
\begin{equation}\label{e:Pobsdef}
P_{\obs}=\frac{2\pi+\Delta \lambda}{n}=P_0\left(1+\frac{\Delta\lambda}{2\pi}\right)
\end{equation}
where $\Delta \lambda=\lambda_{\rm tr}(t_2)-\lambda_{\rm tr}(t_1)$.

In the following we assume that the shift in the mean longitude is caused by the
perihelion shift $\Delta \omega = P_0 \dot\omega$
per period, (i.e. we assume a constant eccentricity), then from the chain rule
\begin{equation}\label{e:DLambda}
 \Delta \lambda= \frac{\partial\lambda}{\partial \omega}\Delta\omega=
\frac{\partial\lambda}{\partial k} \frac{\partial k}{\partial \omega}\Delta\omega + \frac{\partial\lambda}{\partial h} \frac{\partial h}{\partial \omega}\Delta \omega.
\end{equation}
Here $\partial k/\partial \omega=-e\sin\omega$ and $\partial h/\partial \omega=e\cos\omega$, from definition (see above), and the partial derivatives $\partial\lambda/\partial k$ and
$\partial\lambda/\partial h$ can be found from
\eqref{lambdaattransit} and are given explicitly in the
Appendix (\ref{e:lambda_k}--\ref{e:lambda_h}). At transit, we get
\begin{equation}\label{e:lambda/omega}
\frac{\partial \lambda}{\partial \omega}=
\frac{e^2}{1+\sqrt{1-e^2}}+\sqrt{1-e^2}\frac{e^2+(2+e\sin\omega)e\sin\omega}{(1+e\sin\omega)^2}.
\end{equation}
Combining \eqrefs{e:Pobsdef}{e:lambda/omega} and defining
the secular period of the periastron precession as
$P_{\rm sec}=(2\pi)/\dot\omega$, we get
\begin{equation}
P_{\obs}=P_0+\frac{P_0^2}{2\pi} \frac{\partial\lambda}{\partial\omega} \dot\omega
=P_0+ \frac{\partial\lambda}{\partial\omega}\frac{P_0^2}{P_{\rm sec}}.\label{pobs1}
\end{equation}

Since $\partial\lambda/\partial\omega$ itself is not constant due
to periastron precession,
the observed period between transits slowly changes. Differentiating \eqref{pobs1} with respect to time, we get
\begin{equation}\label{pobs1b}
\dot P_{\rm obs} = 2\pi\frac{\partial^2\lambda}{\partial\omega^2} \frac{P_0^2}{P_{\rm sec}^2}
=\frac{4\pi(1-e^2)^{3/2}e\cos\omega}{(1+e\sin\omega)^3}\frac{P_0^2}{P_{\rm sec}^2},
\end{equation}
since the partial derivative of \eqref{e:lambda/omega} with respect to $\omega$ is
\begin{equation}
\frac{\partial^2\lambda}{\partial\omega^2}=\frac{2(1-e^2)^{3/2}e\cos\omega}{(1+e\sin\omega)^3}. \label{d2ldo2}
\end{equation}
Note that this equation clearly shows that the small eccentricity
approximation $\partial^2\lambda/\partial\omega^2 \approx e\cos \omega$ 
\citep[e.g. used by][]{miralda2002,heyl2007} is very imprecise
for moderate to large eccentricites. Depending on the actual
value of $\omega$, \eqref{d2ldo2} can result even $6-8$ times
smaller or larger values for $\partial^2\lambda/\partial\omega^2$
as its first order approximation for eccentricities $0.5-0.7$.

We have to mention here that the observed period and therefore the individual
transit timings are also affected by the light time effect (LTE). Since
the precession of an elliptical orbit causes the distance between
the host star and the planet at the transit instances to vary,
the light time delay will change for transit event to transit event.
The magnitude of this effect can be derived as follows.
The distance between the star and the planet at transit time
(i.e. when $\lambda=\varphi_0=\pi/2$) is
\begin{equation}
r=\frac{a(1-e^2)}{1+e\cos v}=\frac{a(1-e^2)}{1+e\cos(\varphi_0-\omega)}=
\frac{a(1-e^2)}{1+e\sin\omega}. \label{rattransit}
\end{equation}
This difference in the distance implies an additional $-r(1-\mu)/c$ time shift, respective to
the barycentric reference frame of the star--planet system. Here,
$\mu=M_{\rm p}/(M_{\rm p}+M_\star) \ll 1$ is the mass parameter, $a$
is the semimajor axis of the orbit and $c$
is the speed of light. Therefore, the correction in the observed period is
\begin{eqnarray}
P^{\rm +LTE}_{\obs} & = & P_{\obs}-\frac{a}{c}(1-e^2)\frac{\partial (1+e\sin\omega)^{-1}}{\partial\omega}P_0\dot\omega= \nonumber \\
& = & P_{\obs}+2\pi \frac{a}{c}\frac{(1-e^2)e\cos\omega}{(1+e\sin\omega)^2}\frac{P_0}{P_{\rm sec}},
\end{eqnarray}
where we neglected the barycentric correction.
Thus, the variation in this period (corrected for LTE) due to the
variation in $\omega$ is
\begin{equation}\label{e:LC}
\dot P^{\rm +LTE}_{\obs} = \dot P_{\obs} - 4\pi^2(1-e^2)\frac{e(e+e\cos^2\omega+\sin\omega)}{(1+e\sin\omega)^3}\frac{a}{c}\frac{P_0}{P_{\rm sec}^2}.
\end{equation}
Comparing \eqref{pobs1b} and (\ref{e:LC}), and assuming that
the motion of the planet is non-relativistic, $a/c\ll P_0$, we find that
$|\dot P^{\rm +LTE}_{\obs}-\dot P_{\obs}| \ll |\dot P_{\obs}|$.
We conclude that the period variation due to the LTE
is negligible compared to the period variation caused by the changing geometry.

In summary, \eqrefs{pobs1}{pobs1b}, along with \eqref{e:lambda/omega}, give the
modulation of the actual observable period between transit events. These results are valid for arbitrary eccentricities
and are independent of the physical mechanism causing the secular precession of
the periastron. We calculate the secular precession period caused by general relativity \citep[see e.g.,][]{wald1984} using
\begin{equation}
P_{\rm sec} = \frac{c^2 (1-e^2)}{3(2\pi G M_\star)^{2/3}}
P_0^{5/3}, \label{pprecmmotion}
\end{equation}
where $M_\star$ is the mass of the parent star, $G$ is Newton's gravitational
constant. This secular period
is of order $10^4$--$10^5\,$years for the known ETEP systems
(see Table~\ref{table:basicdata} and \scref{s:realsystems}
for more details). We note that if other planets are also present in these systems, they might also cause additional periastron precession of a larger magnitude. The Yarkovski-effect \citep{fabrycky2008} and the tidal circularization \citep[see][and the references therein]{johnskrull2007} lead to negligible modifications for our purposes, as these effects are relevant on timescales exceeding $0.1\,$Gyr for these systems. In the following, we compare the precession measurement sensitivities with the general relativistic rate $P_{\rm sec}$ given by \eqref{pprecmmotion}.

\subsection{Modulation of the transit duration}

Here we investigate how the periastron precession
affects the duration of a transit. Let us denote the half duration of the
transit by $H=\TD/2$, which we define as half the time between the instances when
the center of the planetary disk intersects the limb of the star, i.e. between
the center of the ingress and egress. Note that this
is not the time between the first and last contact.
This is important because the instances of the center of the ingress
and egress can be measured more accurately than their beginning or end.
Since we are interested in estimating the \emph{variations} of the duration
of the transit to leading order, we perform a first-order calculation,
i.e. assuming a constant apparent tangential velocity for the planet and neglecting
the changes in the impact parameter due to the elliptical orbit and/or
the curvature of the projection of the orbit due to the inclination. From 
Kepler's Second Law and \eqref{rattransit}, one can calculate
the tangential distance $\Delta x$  traveled by the planet
during time $H$,
\begin{equation}
\Delta x=v_{\rm tan} H = an\frac{1+e\sin\omega}{\sqrt{1-e^2}} H
\end{equation}
(see \apref{appendixtangentvelocity} for the derivation of $v_{\rm tan}$).
This can be related to the impact parameter $b$ and
the radius of the star $R_\star$ for the geometry of
the transit as
\begin{equation}
\frac{\Delta x}{R_\star}= \sqrt{1-b^2}.
\end{equation}
Thus, to leading order,
\begin{equation}\label{e:H}
H  =  \frac{P_0}{2\pi}\left\{\frac{R_\star}{a}\frac{\sqrt{1-e^2}}{1+e\sin\omega}\sqrt{1-b^2} +
\mathcal{O}\left[\left(\frac{R_\star}{a}\right)^{3}\right]\right\}.
\end{equation}
Note that \eqref{e:H} depends on $\omega$ through the $(1+e\sin\omega)^{-1}$ term and
also implicitly through $b$,
\begin{equation}\label{e:b}
b=\frac{r\cos i}{R_\star} = \frac{a}{R_\star}\frac{1-e^2}{1+e\sin\omega}\cos i.
\end{equation}
The variation in $H$ caused by the variation in the periastron can be found from
\eqref{e:H} and (\ref{e:b}),
\begin{equation}
\frac{\partial H}{\partial\omega}=
\frac{e\cos\omega}{1+e\sin\omega}H\frac{1-2b^2}{1-b^2}. \label{dhdomega}
\end{equation}
Note that equation reflects the qualitative expections implied by
Kepler's Second Law. Namely, if an eccentric orbit advances, the
distance between the planet and the star will change. If this distance
decreases, the impact parameter will also decrease (resulting a longer
transit duration) but due to Kepler's Second Law, the apparent tangential
velocity of the transiting object will increase (resulting a shorter
transit duration). Therefore at a certain value of the inclination
and/or the impact parameter, the two effects cancel each other yielding
no TDV effect. Equation~\ref{dhdomega}
clearly shows that it occurs when the impact parameter is
$b=1/\sqrt{2}\approx0.707$.
The long-term variation in the duration of the transit is then
\begin{equation}\label{doth}
\dot H =\frac{\partial H}{\partial\omega}\dot\omega=\frac{\sqrt{1-e^2} e \cos \omega}{(1+e\sin \omega)^2}\frac{1-2b^2}{\sqrt{1-b^2}}\frac{R_{\star}}{a}\frac{P_0}{P_{\rm sec}}
\end{equation}
Comparing \eqref{pobs1b} and \eqref{doth} the TDV effect relates to the TTV effect as
\begin{equation}\label{e:TDV/TTV}
 \frac{\dot H}{\dot P_{\rm obs}} =
\frac{1+e\sin \omega}{6\pi}\frac{1-2b^2}{\sqrt{1-b^2}} \frac{R_{\star}}{R_{\rm Sch}}
\end{equation}
where $R_{\rm Sch}=2 G M/c^2$ is the Schwarzschild radius of the star. As an example, note that $R_{\star}/R_{\rm Sch}=2.5\times 10^{5}$ for the Sun.
Therefore, \eqref{e:TDV/TTV} shows that the TDV effect is always much larger than the TTV effect. In particular, the ratio is larger for increasing $b$. In the
limit $b\to 1$ \eqref{doth} breaks down because the periastron precession shifts the orbit out of the transiting region.

\subsection{Observational implications}

Now, using the results for the TDV and TTV effects, \eqref{pobs1b} and (\ref{doth}), we can estimate
how these timing and transit duration variations might be observed on
long timescales. In the following we discuss these observational implications.

\subsubsection{Transit Timing Variations}
Equation~(\ref{pobs1b}) shows that the observed period between successive transits increases
or decreases at a practically constant rate  during the observations, $\dot P_{\obs}$.
The time of the $m$th transit from an arbitrary epoch $T_0$ can be found from adding up the contributions of
the observed $m$ number of periods
\begin{eqnarray}
T_m&=& T_0 + P_{\obs}m+Dm^2, \label{ttiming}
\end{eqnarray}
where $P_{\obs}\approx P_0$ denotes the time of the first observed orbit,
$D$ is the transit timing variation factor,
\begin{equation}
D=\frac{P_{0}\dot P_{\obs}}{2}=2\pi G_{\rm TV}(e,\omega)P_0\left(\frac{P_0}{P_{\rm sec}}\right)^2, \label{dtiming}
\end{equation}
where we have introduced the geometrical factor
\begin{equation}\label{e:G_TV}
G_{\rm TV}(e,\omega)=(1-e^2)^{3/2}\frac{e\cos\omega}{(1+e\sin\omega)^3}.
\end{equation}

The chance of detecting the periastron precession increases
with the geometrical factor $G_{\rm TV}(e,\omega)$ which is related to
the alignment of the orbital ellipse with the line of sight.
The optimal value for detecting the precession is $\omega=0$ or $\pi$
for small eccentricities, i.e. the semimajor axis
should be perpendicular to the line of sight. For arbitrary eccentricities,
the optimal value for $\omega$ for the TTV observation is
\begin{equation}
\omega^{\rm best}_{\rm TV}=\frac{3}{2}\pi\pm\arccos\left(\frac{6e}{1+\sqrt{1+24e^2}}\right).
\end{equation}
which approaches $0^{\circ}$ and $180^{\circ}$ for small eccentricities as expected,
and $270^{\circ}$ for large eccentricities. In case of $e=0.5$,
$\omega_{\rm extr}=\{235.4^\circ,304.6^\circ\}$. The least favorable value for $\omega$ occurs
when $G_{\rm TV}(e,\omega)=0$, i.e. at $\omega^{\rm worst}_{\rm TV}=\{90^{\circ},270^{\circ}\}$,.

\subsubsection{Transit Duration Variations}
Now let us turn to the TDV effect. The observed duration of the $m$th transit can be calculated in the same way,
namely
\begin{equation}
H_m=H_0+Fm,
\end{equation}
where $F$ is the shift in the transit duration per orbit. This factor is
\begin{equation}\label{e:F}
 F= P_0 \dot H = 2\pi G_{\rm DV}(e,\omega,b)H\frac{P_0}{P_{\rm sec}},
\end{equation}
and
\begin{equation}\label{e:G_DV}
 G_{\rm DV}(e,\omega,b)=\frac{1-2b^2}{1-b^2}\frac{e\cos\omega}{1+e\sin\omega}.
\end{equation}
The optimal orientation $\omega^{\rm best}_{\rm DV}$ for detecting the TDV effect for fixed $e$, $b$, and $P_0$ can be found by maximizing $ H G_{\rm DV}(e,\omega,b)$. The result is
\begin{equation}
\omega^{\rm best}_{\rm DV}=\frac{3}{2}\pi\pm\arccos\left(\frac{4e}{1+\sqrt{1+8e^2}}\right).
\end{equation}
and the worst orientation is at
$\omega^{\rm worst}_{\rm DV}=\{90^{\circ},270^{\circ}\}$,
just like for the TTV case. Comparing $\omega^{\rm best}_{\rm TV}$ and
$\omega^{\rm best}_{\rm DV}$ it is clear that the most favorable
orientation in terms of the two effects are similar, hence
the chance of detecting the periastron motion through
transit timing variations or transit duration variations
is correlated. Both effects go away if the eccentricity is oriented
parallel to the line of sight.
We also note that for moderate values of $e$ and small impact parameters,
$|G_{\rm TV}|\approx|G_{\rm DV}|$ which also implies that the
most favorable geometry for detecting either TTVs or TDVs is similar.

\subsection{Error analysis}
Next we estimate the parameter measurement
precision of the TTV and TDV effects for future observations.
We consider the repeated observation
of a particular transiting system over a total timespan $T_{\rm tot}$,
measuring the transit timing $T_m$ and duration $H_m$ for each transit
with respective errors $\sigma(T)$ and $\sigma(H)$. (We discuss the
specific values of $\sigma(T)$ and $\sigma(H)$ for transit observations
in Section~\ref{s:photometric}). For simplicity,
let us assume that these measurements are equidistant and in total
$N$ independent transits are observed, i.e. the $m$th transit is observed if
$m=0,d,\dots,(N-1)d$, where $d=T_{\rm tot}/(NP_0)$.

\subsubsection{Transit Timing Variations}
Using \eqref{ttiming}, we can fit a second-order polynomial to these
observations with unknown coefficients $T_0$, $P_{\rm obs}$ and
$D$ by minimizing the merit function
\begin{equation}
\chi^2_{\rm TV} = \sum\limits_{m=0,d,\dots,(N-1)d}
\left[\frac{T_m - (T_0 + P_{\rm obs}m+Dm^2)}{\sigma(T)}\right]^2.
\end{equation}
The minimization of the above function results in a linear set of equations
in the parameters $p_{i}=\{T_0, P_{\rm obs}, D\}$. Assuming Gaussian errors,
the parameter estimation covariance matrix can be found from the
Fisher matrix method \citep{finn1992}:
\begin{equation}
 \left< \delta p_i \delta p_j \right> = (\mathcal{F}^{-1})_{ij}
\end{equation}
Here $\mathcal{F}$ is the Fisher matrix defined as
\begin{equation}
\mathcal{F}_{ij} = \sum\limits_{m=0,d,\dots,(N-1)d} \frac{1}{\sigma^2(T)}\frac{\partial T^{\rm fid}_m}{\partial p_i} \frac{\partial T^{\rm fid}_m}{\partial p_j}
\end{equation}
where $T^{\rm fid}_m$ is the fiducial value of $T_m$ given by \eqref{ttiming}.
The marginalized expected squared parameter estimation error is given by the diagonal elements of the covariance error matrix $\sigma^2(p_i)=(\mathcal{F}^{-1})_{ii}$. In particular,
the resulting uncertainty of $D$ becomes
\begin{eqnarray}
\sigma(D) &=& \frac{\sqrt{180}\,\sigma(T)}{d^2 \sqrt{N(N^2-1)(N^2-4)}} \nonumber\\
&\approx& \sqrt{180} \left(\frac{P_0}{T_{\rm tot}}\right)^2 \left(1+\frac{5}{2N^2} \right) \frac{\sigma(T)}{\sqrt{N}}. \label{duncert}
\end{eqnarray}
Here, the first equality is valid for arbitrary $N$, while the second
is its first order approximation for large $N$. The leading order
approximation is verified against \citet{press1992}.

\subsubsection{Transit Duration Variations}
We can repeat the same calculations as above for the observation of the
half transit duration $H_m$ to measure the variation factor $F$.
The merit function in this case
\begin{equation}
\chi^2_{\rm DV} = \sum\limits_{m=0,d,\dots,(N-1)d}
\left[\frac{H_m - (H_0 + Fm)}{\sigma(H)}\right]^2,
\end{equation}
has to be minimized for the same set of observations.
This minimization again leads to a linear set of equations in the
parameters $H_0$ and $F$. The Fisher matrix method in this case
gives the uncertainty in $F$ as
\begin{eqnarray}
\sigma(F) &=& \frac{\sqrt{12}\sigma(H)}{d\sqrt{N(N^2-1)}}\approx \nonumber\\
&\approx& \sqrt{12} \frac{P_0}{T_{\rm tot}} \left(1+\frac{1}{2N^2} \right)
\frac{\sigma(H)}{\sqrt{N}}. \label{funcert}
\end{eqnarray}

\begin{figure}
\resizebox{8cm}{!}{\includegraphics{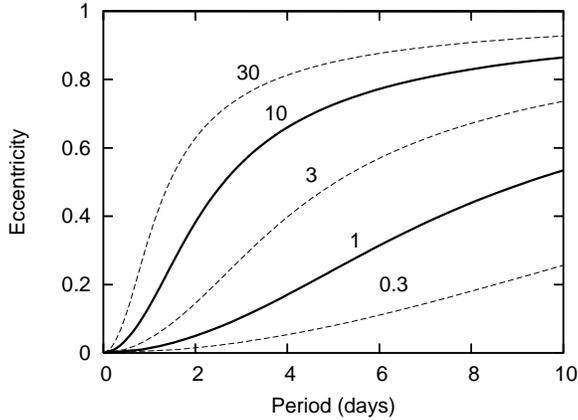}}
\caption{%
The significance of detecting the GR periastron precession $|F_0|/\sigma_0(F)$
through the TDV effect as a function
of orbital period and eccentricity. The transit duration is assumed to be measured for $4$~years each with a $5$~sec error,
for a Sun-like star on a non-inclined orbit.
Increasing the mass or radius of the star, or the impact parameter increases the
detection significance (see text).
}\label{fig:tdvksig}
\end{figure}
Figure~\ref{fig:tdvksig} shows the detection significance of the TDV
measurement $|F|/\sigma(F)$, if the precession rate in $F$ is given by
the general relativistic formula, \eqref{e:F}.
Here each transit is assumed to be measured (i.e. $d=1$) with a precision $\sigma(H)=5$\,sec
for a total observation time of $T_{\rm tot}=4$\,years. These assumptions are
realistic for the future Kepler mission (see \S~\ref{s:photometric} below).
Other parameters are $M_{\star}=M_{\odot}$, $R_{\star}=R_{\odot}$, $b=0$, and
we averaged over the possible orientations of $\omega$.
For other parameters,
\begin{eqnarray}
\frac{|F|/\sigma(F)}{|F_{0}|/\sigma(F_{0})} & = &  \frac{1}{d}\frac{1}{\sqrt{1-b^2}}
\left(\frac{R_{\star}}{R_{\odot}}\right) \left(\frac{M_{\star}}{M_{\odot}}\right)^{1/3}\cdot \nonumber \\
& & \cdot
\left(\frac{\sigma(H)}{5\,{\rm sec}}\right)^{-1}
\left(\frac{T_{\rm tot}}{4\,{\rm yr}}\right)^{3/2},
\end{eqnarray}
implying that the detection significance can be even better.

Figure~\ref{fig:tdvksig} clearly shows that the chances of detecting the
precession effects
through the TDV effect is encouraging. The detection significance of the
general relativistic precession of a transiting
exoplanet with eccentricity $e\gtrsim 0.2$ and period $P\lesssim 5$~days, is
typically over the \sigmalevel{1} level.
Generally, \eqref{duncert} and \eqref{funcert} can be used directly to
check what kind of observations are required to detect
the precession of the periastron through the TTV or TDV methods, respectively.


\section{The case of XO-3\lowercase{b},
HD~147506\lowercase{b}, GJ436\lowercase{b} and HD~17156\lowercase{b}}
\label{s:realsystems}

As of this writing, four TEPs are known with a non-zero eccentricity within
\sigmalevel{3}, namely
XO-3\lowercase{b} \citep{johnskrull2007},
HD~147506\lowercase{b} \citep{bakos2007},
GJ~436b\lowercase{b} \citep{butler2004,gillon2007},
and HD~17156b\lowercase{b} \citep{fischer2007,barbieri2007}.
The planet TrES-1 \citep{alonso2004} has also been reported as an object with
non-zero eccentricity, however, it is zero within \sigmalevel{2} thus we
omit from our analysis. We note here that recently both GJ~436
and HD~17156 have been suggested to have another planetary companions
\citep[see][]{ribas2008,short2008}.
The secular period of the periastron motion
are determined by the mass of the star $M_\star$, the orbital period $P_0$,
and the eccentricity $e$,
while the timing variation constant $D$ is also affected by the
actual argument of pericenter, $\omega$. The transit duration variation
factor $F$ is affected indirectly by the geometrical ratio
$a/R_\star$ and directly by the impact parameter $b$.
These parameters are summarized in the first seven columns of Table~\ref{table:basicdata}
for these four ETEP systems. The derived GR periastron precession period,
$P_{\rm sec}$ can be found in the 8th column of the table. 

In addition to the inevitable periastron precession caused by GR, there
might be other sources of perturbations causing periastron precession.
The last column gives the minimum mass to semimajor axis ratio
of a hypothetical exterior perturber (e.g. a planet or an asteroid belt), in Earth mass units, which
causes the same periastron presession rate as that caused by the general
relativity. This estimate based on \citet{price1979}, and is valid for
$a_2\gtrsim 3a$ and for non-resonant cases.
Note that the minimum mass of the perturber scales with the third power of the
semimajor axis ratio to cause a comparable precession rate as GR.
The numbers show that the precession caused by additional planets
in the system, if present, can easily cause a larger precession rate than GR.
In case of orbital resonances with exterior planets, 
the precession rate can be even larger \citep{holman2005,agol2005}.
To be conservative, we examine whether the precession rate can be measured
to a precision better than that corresponding to GR.

To obtain a high significance, \sigmalevel{k} detection of the
periastron precession using
transit timing variations, we need $k\sigma(D)\approx|D|$.
Using \eqref{duncert},
the total number of transits necessary to measure $D$ with this precision is
\begin{equation}
N_{\rm TV}=180\left[\frac{\sigma(T)}{k|D|}\right]^2\left(\frac{P_0}{T_{\rm tot}}\right)^4.
\end{equation}
Thus, the number of such required transit observations
is extremely sensitive to the orbital period $P_0$ and the observation timespan.
Table~\ref{table:ttvdata} gives the corresponding values of
the transit timing variation factors for the known ETEP systems
and this number of observations, assuming a $T_{\rm tot}=20$~year long
observational timespan, timing precision of $\sigma(T)=2~{\rm sec}$
and \sigmalevel{3} sensitivity of the GR periastron precession
level. The table shows the recently discovered XO-3b system is a promising
candidate to detect the GR periastron precession through the TTV effect,
while the other ETEP systems require unrealistically many observations. We note
that if other perturbing planets are present in these systems and lead to a
precession rate that is larger by a factor of 10 than the GR precession rate,
then the number of detections (during the same $T_{\rm tot}$ timespan) is
lower by a factor of 100. It is interesting that the best candidate (by far) is
XO-3b even though its eccentricity is not as large as that of
HD~147506b or HD~17156b.


\begin{table}
\caption{Transit timing variation factor ($D$, in seconds) and the
number of transits
($N_{20\,{\rm y},\sigmalevel{3},2\,{\rm sec}}$) what should be
detected almost uniformly in a 20 year long timespan, each with an error of
$2$ second to confirm the precession within \sigmalevel{3}.
}\label{table:ttvdata}
\begin{center}\begin{tabular}{llr}
\hline
System	 	& $D$ (seconds)			
		& $N_{20\,{\rm y},~\sigmalevel{3},~2\,{\rm sec}}$	\\
\hline
HD~147506b	& $-(5.9\pm0.7)\cdot10^{-7}$ 	
		& $6600$			\\
XO-3b		& $+(4.3\pm0.6)\cdot10^{-7}$
		& $1280$			\\
GJ~436b		& $+(5.0\pm0.5)\cdot10^{-8}$  	
		& $44160$			\\
HD~17156b	& $-(6.9\pm0.8)\cdot10^{-8}$ 	
		& $97\cdot10^{6}$		\\
\hline
\end{tabular}\end{center}
\end{table}

\begin{table}
\caption{Transit duration variation factor ($F$, in seconds) and the
number of transits
($N_{2\,{\rm y},~\sigmalevel{3},~2\,{\rm sec}}$) what should be
detected almost uniformly in a 4 year long timespan, each with an error of
$2$\,sec to confirm the precession within \sigmalevel{3}.
}\label{table:tdvdata}
\begin{center}\begin{tabular}{llr}
\hline
System	 	& $F$ (seconds)			
		& $N_{4\,{\rm y},~\sigmalevel{3},~2\,{\rm sec}}$	\\
\hline
HD~147506b	& $-(1.8\pm0.3)\cdot10^{-2}$ 	
		& $20$				\\
XO-3b		& $-(5.4\pm0.6)\cdot10^{-3}$
		& $70$				\\
GJ~436b		& $-(4.1\pm0.5)\cdot10^{-3}$  	
		& $85$				\\
HD~17156b	& $-(3.2\pm0.2)\cdot10^{-3}$ 	
		& $8970$		\\
\hline
\end{tabular}\end{center}
\end{table}


Let us now turn to the observational constraints for the detection
of transit duration variations. Using \eqref{funcert},
the total number of required observations within $T_{\rm tot}$ is
\begin{equation}
N_{\rm DV}=12\left[\frac{\sigma(H)}{k|F|}\right]^2
\left(\frac{P_0}{T_{\rm tot}}\right)^2.
\end{equation}
Note that $N_{\rm DV}$ is not as sensitive to the $P_0/T_{\rm tot}$
ratio as $N_{\rm TV}$, and implies that a smaller number of observations is
typically necessary.

In Table~\ref{table:tdvdata} we present the values of the transit
duration variation factor, $F$, and the number of required observations
to reach the same \sigmalevel{3} confidence for detecting the GR
periastron precession\footnote{Note that since the measurement of the
TDV effect relies on fitting 2 parameters, instead of 3 parameters
for the TTV effect, the \sigmalevel{3} confidence corresponds to a higher
confidence level for the TDV effect.}.
Here we assumed a shorter observation timespan,
$T_{\rm tot}=4$~year (i.e. shorter compared to the 20~year long timespan
necessary for the detection of the TTV effect),
the same timing precision
of $\sigma(H)=2~{\rm sec}$ and the same level of detection, \sigmalevel{3}.
The best known ETEP system for TDV detection is therefore HD~147506b,
but the number of necessary observations
is feasible for XO-3b and GJ~436b as well.

\subsection{Photometric detection}
\label{s:photometric}

The precision for measuring the transit timing and transit duration
for a photometric observation can be estimated as follows. Since
the time of the ingress ($T_{\rm I}$) and the time of the egress
($T_{\rm E}$) -- i.e. when the center of the planet crosses the
limb of the star inwards or outwards, respectively --
defines both the time of the transit center and
the half duration like
\begin{eqnarray}
T & = & \frac12(T_{\rm E}+T_{\rm I}), \\
H & = & \frac12(T_{\rm E}-T_{\rm I}),
\end{eqnarray}
moreover $T_{\rm I}$ and $T_{\rm E}$ can be treated as uncorrelated variables,
therefore the uncertainties of the transit time and half duration would
be nearly the same, i.e. $\sigma(T)\approx\sigma(H)$. We have estimated
these uncertainties for the four distinct planets
using Monte-Carlo simulations by fitting
transit light curves on mock data sets. We have used the
observed planetary parameters as an input for these artificial light curves.
The fit was performed assuming
quadratic limb darkening \citep[see][]{mandel2002} in the Sloan $z'$ band.
The mock light curves were sampled with $\Delta\tau_1=1\,{\rm sec}$
cadence and an additional Gaussian noise of $\sigma_1(m)=1\,{\rm mmag}$
was added. The resulting uncertainties, $\sigma_{1,1}(T)$ and $\sigma_{1,1}(H)$
for the four planets are presented in Table~\ref{table:ttvtdvuncert}. Since
the depth of the four transits are nearly the same (see the appropriate
normalized radii, $p=R_p/R_\star$, all between $0.068\lesssim p\lesssim0.085$),
the uncertainties $\sigma_{1,1}(T)$ and $\sigma_{1,1}(H)$ are almost the same
for the four cases. Using these normalized values, one can
easily estimate the uncertainties for arbitrary sampling cadence
$\Delta \tau$ and photometric precision $\sigma(m)$ using
\begin{eqnarray}
\sigma(T) & \approx & \sigma_{1,1}(T)\frac{\sigma(m)}{1\,\rm mmag}
\sqrt{\frac{\Delta\tau}{1\,\rm sec}}, \\
\sigma(H) & \approx & \sigma_{1,1}(H)\frac{\sigma(m)}{1\,\rm mmag}
\sqrt{\frac{\Delta\tau}{1\,\rm sec}}.
\end{eqnarray}
For comparison, note that the expected photometric precision of the Kepler
space telescope \citep[see][]{borucki2007} is $1\,{\rm mmag}$
for observing a light curve of a bright, $M_v=8.8$ star with a
$1\,{\rm sec}$ sampling cadence. Since
the star XO-3 has almost the same apparent magnitude, it is clear,
the transit durations
would be detected with an accuracy of $\sigma(H)\approx 5\,{\rm sec}$
if this star was in the field of Kepler.
Therefore, \eqref{funcert} and Table~\ref{table:tdvdata} shows
that the transit duration variations
would be detectable for HD~147506(b) or XO-3(b)--like systems
due to the GR periastron precession,
within \sigmalevel{3} confidence
with a Kepler--type mission within approximately 3 or 4 years,
respectively.


\begin{table}
\caption{Uncertainties of the transit time and transit duration measurements
for the four known ETEPs, assuming Sloan $z'$-band photometric data
taken with a 1\,sec cadence and 1\,mmag photometric precision.
}\label{table:ttvtdvuncert}
\begin{center}\begin{tabular}{lll}
\hline
System	 	& $\sigma_{1,1}(T)$ (sec)			
		& $\sigma_{1,1}(H)$	(sec) \\
\hline
HD~147506b	& 5.3	& 4.8	\\
XO-3b		& 6.9	& 4.7	\\
GJ~436b		& 6.7	& 8.4	\\
HD~17156b	& 5.4	& 6.1	\\
\hline
\end{tabular}\end{center}
\end{table}



\section{Summary}
\label{s:summary}

The first four eccentric transiting exoplanetary
systems have been discovered during 2007. The precession of an eccentric orbit
causes variations both in the transit timings and transit durations.
We estimated the significance of measuring the corresponding observable effects
compared to the inevitable precession rate of general relativity.
We applied these
calculations to predict the significance of measuring the
effect for the four known eccentric transiting planetary systems.
Our calculations show that a space-borne telescope is
adequate to detect the change in the transit durations
to a high significance better than the GR periastron precession rate
within a 3~--~4 year timespan (in a continuous observing mode).
The same kind of instruments would need more than a decade
to detect the corresponding transit time variations to this sensitivity
even for the most optimistic known system.

The CoRoT mission has already found two transiting planets
\citep[see][]{barge2008,alonso2008} and there are two known
planets in the planned field-of-view of the Kepler mission
\citep[see][]{odonovan2006,pal2008}.
Our results suggest that if an \emph{eccentric} transiting planet
is found in the Kepler or CoRoT field, these missions
will be able to measure the periastron
precession rate to a very high significance within their mission lifetime
or with the support of ground-based observations on a longer time scale.
This will provide an independent test of the theory of
general relativity and will also be useful for testing for the
presence of other planets in these systems.

\section*{Acknowledgments}

The authors would like to thank the hospitality and support of
the Harvard-Smithsonian Center for Astrophysics where this
work was partially carried out.
We thank Andres Jordan for useful comments on the manuscript.
BK acknowledges support from OTKA grant No.~68228.


{}


\appendix
\section{Mean longitude at the transit instances}
\label{appendixlambdaderiv}

The derivation of \eqref{lambdaattransit} goes as follows. According to
Kepler's equation, $E-e\sin E=M=\lambda-\omega$, one can write
$\lambda=\omega+E-e\sin E$. The only thing what is to be done is to
calculate the eccentric anomaly $E$ for the instance when the orbiting
body intersect the semi-line with the argument angle $\varphi_0$. The
latter means that the true anomaly $v$ of the body is $v=\varphi_0-\omega$,
by definition. The relation between the eccentric and true anomaly
is
\begin{equation}
\tan\frac{E}{2}=\sqrt{\frac{1-e}{1+e}}\tan{\frac{v}{2}},
\end{equation}
which is equivalent with
\begin{eqnarray}
\cos E & = & \frac{e+\cos v}{1+e\cos v}, \label{coseatt}\\
\sin E & = & \frac{\sqrt{1-e^2}\sin v}{1+e\cos v}. \label{sineatt}
\end{eqnarray}
Using the addition theorem, the sine and cosine of the angle $\omega+E$
can be written as:
\begin{eqnarray}
\cos(\omega+E) & = & \cos\omega\frac{e+\cos(\varphi_0-\omega)}{1+e\cos(\varphi_0-\omega)} - \nonumber \\
	& & - \sin\omega\frac{\sqrt{1-e^2}\sin(\varphi_0-\omega)}{1+e\cos(\varphi_0-\omega)}, \\
\sin(\omega+E) & = & \sin\omega\frac{e+\cos(\varphi_0-\omega)}{1+e\cos(\varphi_0-\omega)} + \nonumber \\
	& & + \cos\omega\frac{\sqrt{1-e^2}\sin(\varphi_0-\omega)}{1+e\cos(\varphi_0-\omega)}.
\end{eqnarray}
Thus, the mean longitude itself is going to be
\begin{eqnarray}
\lambda & = & \omega+E-e\sin E=\arg\left[\cos(\omega+E),\sin(\omega+E)\right]- \nonumber \\
	& & - e\frac{\sqrt{1-e^2}\sin v}{1+e\cos v}.
\end{eqnarray}
If both arguments of the above $\arg[\cdot,\cdot]$ function is
multiplied by the always positive common denominator
$1+e\cos(\varphi_0-\omega)$, one gets after some simplification:
\begin{eqnarray}
\omega+E & = & \arg\left[k+\cos\varphi_0+\frac{h(k\sin\varphi_0-h\cos\varphi_0)}{1+\sqrt{1-e^2}}, \right. \nonumber \\
	& & \left.h+\sin\varphi_0-\frac{k(k\sin\varphi_0-h\cos\varphi_0)}{1+\sqrt{1-e^2}}\right].
\end{eqnarray}
Putting all terms together and replacing the appropriate terms by
$e_\perp=k\sin\varphi_0-h\cos\varphi_0$,
$e_\parallel=k\cos\varphi_0+h\sin\varphi_0$ and $\ell=1-\sqrt{1-e^2}$,
we get \eqref{lambdaattransit}.
The partial derivatives of \eqref{lambdaattransit} become
\begin{eqnarray}
\frac{\partial\lambda}{\partial k} & = & -\frac{h}{2-\ell}-(1-\ell)\frac{h+(2+e_
\parallel)\sin\varphi_0}{(1+e_\parallel)^2}, \label{e:lambda_k}\\
\frac{\partial\lambda}{\partial h} & = & +\frac{k}{2-\ell}+(1-\ell)\frac{k+(2+e_
\parallel)\cos\varphi_0}{(1+e_\parallel)^2}. \label{e:lambda_h}
\end{eqnarray}

\section{Tangential velocity and position at the transit}
\label{appendixtangentvelocity}

It is known from the theory of the two-body problem that the
angular momentum of a body orbiting around a mass of $GM=\mu$
and having an orbit with the semimajor axis of $a$ and eccentricity $e$
is $C=\sqrt{\mu a(1-e^2)}$. Since $C=rv_{\rm tan}$ for all points,
the tangential velocity would be
\begin{equation}
v_{\rm tan}=\frac{C}{r}=\sqrt{\mu a(1-e^2)}\frac{1+e\cos(\varphi_0-\omega)}{a(1-e^2)}\label{vtran1}
\end{equation}
Using Kepler's Third Law, i.e. $\mu=n^2a^3$, the above equation can be
reordered to
\begin{equation}
v_{\rm tan}=an\frac{1+e\cos(\varphi_0-\omega)}{\sqrt{1-e^2}}.\label{vtran2}
\end{equation}
For $\varphi_0=\pi/2$, \eqref{vtran2} becomes
\begin{equation}
v_{\rm tan}=an\frac{1+e\sin\omega}{\sqrt{1-e^2}}.
\end{equation}

\label{lastpage}

\end{document}